\documentclass[prd,showpacs,showkeys,floatfix,nofootinbib,
               preprint,12pt,tightenlines,fleqn]{revtex4}

\usepackage{amsmath,amssymb,graphicx,dcolumn}

\newcommand{\dd}{\mathrm{d}}                    
\newcommand{\id}{\mathrm{d}}                    
\newcommand{\ii}{\mathrm{i}}                    
\newcommand{\action}{\mathcal{S}}               
\newcommand{\BigO}{\mathrm{O}}                  
\newcommand{\diag}{\ensuremath{\mathrm{diag}}}  
\newcommand{\vecbf}[1]{\boldsymbol{\mathrm{#1}}}
\newcommand{\modMaxth}{modified-Max\-well theory} %

\begin{document}

\noindent Phys. Rev. D 77, 016002 (2008)
\hfill arXiv:0709.2502v6 [hep-ph]\newline\vspace*{0.5cm}          
\title[Ultrahigh-energy cosmic-ray bounds]
      {Ultrahigh-energy cosmic-ray bounds on nonbirefringent
       modified-Maxwell theory\vspace*{.125\baselineskip}}
\author{F.R.~Klinkhamer}
\email{frans.klinkhamer@physik.uni-karlsruhe.de}
\thanks{corresponding author}
\affiliation{\mbox{Institute for Theoretical Physics, University of Karlsruhe (TH),}\\
76128 Karlsruhe, Germany}
\author{M.~Risse}\email{risse@physik.uni-wuppertal.de}
\affiliation{\mbox{University of Wuppertal, Physics Department,}\\
Gau\ss stra\ss e 20, 42097 Wuppertal, Germany\\}

\begin{abstract}
\vspace*{2.5mm}\noindent A particularly simple  Lorentz-violating
modification of the Maxwell theory of photons maintains gauge invariance,
CPT, and renormalization. This modified-Maxwell theory, coupled to standard
Dirac particles, involves nineteen dimensionless ``deformation parameters.''
Ten of these parameters lead to birefringence and are already tightly
constrained by astrophysics. New bounds on the remaining nine nonbirefringent
parameters can be obtained from the absence of vacuum Cherenkov radiation for
ultrahigh-energy cosmic rays (UHECRs). Using selected UHECR events recorded
at the Pierre Auger Observatory and assigning pseudo-random directions (i.e.,
assuming large-scale isotropy), Cherenkov bounds are found at the $10^{-18}$
level, which improve considerably upon current laboratory bounds. Future
UHECR observations may reduce these Cherenkov bounds to the $10^{-23}$
level. \vspace*{0cm}\newline
An Addendum with two-sided bounds
has been published separately [Phys. Rev. D 77, 117901  (2008), arXiv:0806.4351].
\end{abstract}

\maketitle

\section{Introduction}
\label{sec:introduction}

It is possible that entirely new phenomena at
the high-energy frontier of elementary particle physics
(with characteristic energy $E_\text{Planck}$ $\equiv$
$\sqrt{\hbar\, c^5/G}$ $\approx$ $1.2 \times 10^{19}\,\text{GeV}$)
lead to Lorentz-violating effects in the low-energy
theory, in particular, the Maxwell theory of photons.
The simplest CPT--invariant Lorentz-violating modification
of the quadratic Maxwell action
involves nineteen dimensionless parameters
\cite{ChadhaNielsen1982,ColladayKostelecky1998,BaileyKostelecky2004}.

Existing bounds on these nineteen ``deformation parameters''
of \modMaxth~are as follows:
ten birefringent parameters are bounded at the $10^{-32}$ level or better
from astrophysics (see Refs.~\cite{KosteleckyMewes2002,KosteleckyMewes2006}
and references therein) and  nine nonbirefringent parameters are
bounded at the $10^{-16}$ level or worse from laboratory experiments
(see Refs.~\cite{Saathoff-etal2003,Odom-etal2006,Stanwix-etal2006}
and references therein).

Following earlier suggestions \cite{Beall1970,ColemanGlashow1997},
it has been noted \cite{Altschul2007PRL98,KaufholdKlinkhamer2007}
that ultrahigh-energy cosmic rays (UHECRs) have the potential
to place further limits on these nonbirefringent parameters
by the inferred absence of ``vacuum Cherenkov radiation.''
In fact, Ref.~\cite{KaufholdKlinkhamer2007} already
gave a bound at the $10^{-17}$ level
on one nonbirefringent parameter from the existing UHECR data
(cf. Refs.~\cite{BhattacharjeeSigl1998,Stanev2004}).
In this article, we obtain even better bounds for \emph{all}
nine nonbirefringent parameters from selected UHECR events
recorded by the Pierre Auger Observatory \cite{Abraham-etal2004}
over the period January 2004 to February 2006 \cite{Abraham-etal2006}.

The present article is organized as follows.
In Sec.~\ref{sec:theory}, we review the theory
considered, quantum electrodynamics with a modified-Maxwell term,
and establish our notation.
Also, we recall the Cherenkov threshold condition on which
our bounds will be based.
In Sec.~\ref{sec:data-selection}, we discuss the selection of an
appropriate set of UHECR events from two years running of
the partially completed Auger experiment.
In Sec.~\ref{sec:UHECRbounds}, we obtain from the Cherenkov threshold condition
and the selected Auger events (with pseudo-random directions)
the promised bounds
on the nine deformation parameters of nonbirefringent
\modMaxth.
In Sec.~\ref{sec:summary}, we summarize
our findings and compare these new astrophysics bounds with
previous laboratory bounds.

Throughout this article, we employ Cartesian coordinates
$(x^\mu)$ $=$ $(x^0,\boldsymbol{x})$ $=$ $(c\,t,x^1,x^2,x^3)$ and
the Minkowski metric $(\eta_{\mu\nu})$ $=$ $\diag(+1$,$-1$,$-1$,$-1)$.
Indices are lowered with the Minkowski metric $\eta_{\mu\nu}$
and raised with the inverse metric $\eta^{\mu\nu}$.
Repeated upper and lower indices are summed over (Einstein summation convention).
The direction of a 3--vector $\boldsymbol{x}$ is given by the unit
3--vector $\widehat{\boldsymbol{x}}\equiv\boldsymbol{x}/|\boldsymbol{x}|$.
Natural units with $c=\hbar=1$ are used, unless stated otherwise.

\section{Theory}
\label{sec:theory}

\subsection{Nonbirefringent modified-Maxwell theory}
\label{sec:nonbirefringent-modified-Maxwell-theory}

The action of modified-Maxwell theory
\cite{ChadhaNielsen1982,ColladayKostelecky1998,BaileyKostelecky2004}
is given by
\begin{equation}\label{eq:modM-action}
\action_\text{modM} =
\int_{\mathbb{R}^4} \id^4 x \;
\Big( -\textstyle{\frac{1}{4}}\,
\big(
\eta^{\mu\rho}\eta^{\nu\sigma}
+\kappa^{\mu\nu\rho\sigma}\big) \,
F_{\mu\nu}(x)\,F_{\rho\sigma}(x)
\Big)\,,
\end{equation}
where $F_{\mu\nu}(x)\equiv \partial_\mu A_\nu(x)-\partial_\nu A_\mu(x)$
is the standard Maxwell field strength of gauge fields $A_\mu(x)$
propagating over flat Minkowski spacetime with a metric $\eta_{\mu\nu}$
as defined in Sec.~\ref{sec:introduction}.
Here, $\kappa^{\mu\nu\rho\sigma}$ corresponds to
a constant background tensor (with real and dimensionless components)
having the same symmetries as the Riemann curvature tensor and
a double trace condition $\kappa^{\mu\nu}_{\phantom{\mu\nu}\mu\nu}=0$,
so that there are $20-1=19$ independent components.
All components of the $\kappa$--tensor in \eqref{eq:modM-action}
are assumed to be very small, $|\kappa^{\mu\nu\rho\sigma}|\ll 1$,
in order to ensure energy positivity \cite{ColladayKostelecky1998}.
The photonic action term \eqref{eq:modM-action} is
gauge-invariant, CPT-even, and power-counting renormalizable
(cf. Ref.~\cite{ItzyksonZuber1980}).

In order to restrict modified-Maxwell theory to the nonbirefringent
sector, the following \emph{Ansatz} for $\kappa^{\mu\nu\rho\sigma}$
has been suggested \cite{BaileyKostelecky2004,Altschul2007PRL98}:
\begin{equation}\label{eq:nonbirefringent-ansatz}
\kappa^{\mu\nu\rho\sigma} =
\textstyle{\frac{1}{2}} \big(\,
 \eta^{\mu\rho}\,\widetilde{\kappa}^{\nu\sigma}
-\eta^{\nu\rho}\,\widetilde{\kappa}^{\mu\sigma}
+\eta^{\nu\sigma}\,\widetilde{\kappa}^{\mu\rho}
-\eta^{\mu\sigma}\,\widetilde{\kappa}^{\nu\rho}
\,\big) ,
\end{equation}
in terms of the nine components of a
symmetric and traceless matrix $\widetilde{\kappa}^{\mu\nu}$,
\begin{equation}\label{eq:widetilde-kappa}
\widetilde{\kappa}^{\mu\nu}=\widetilde{\kappa}^{\nu\mu}\,,\quad
\widetilde{\kappa}^{\mu}_{\phantom{\mu}\mu}=0\,.
\end{equation}
The nine Lorentz-violating ``deformation parameters''
$\widetilde{\kappa}^{\mu\nu}$,
called ``coupling constants'' in Ref.~\cite{KaufholdKlinkhamer2007},
can be rewritten as follows:
\newcommand{\third}{\textstyle{\frac{1}{3}}}
\begin{equation}\label{eq:widetilde-kappa-mu-nu-Ansatz}
\big(\widetilde{\kappa}^{\mu\nu}\big) \equiv
\overline{\kappa}^{00}\,\text{diag}\big(1,\third,\third,\third\big)
+\big(\delta\widetilde{\kappa}^{\mu\nu}\big),\;\;
\delta\widetilde{\kappa}^{00}=0\,,
\end{equation}
with one independent parameter $\overline{\kappa}^{00}$
for the spatially isotropic part of $\widetilde{\kappa}^{\mu\nu}$
and eight independent parameters in $\delta\widetilde{\kappa}^{\mu\nu}$.
For later use, we already define
\begin{equation}\label{eq:alpha-parameters}
\vec{\alpha} \equiv
\left(
  \begin{array}{c}
    \alpha^0 \\
    \alpha^1 \\
    \alpha^2 \\
    \alpha^3 \\
    \alpha^4 \\
    \alpha^5 \\
    \alpha^6 \\
    \alpha^7 \\
    \alpha^8 \\
  \end{array}
\right)
 \equiv
\left(
  \begin{array}{c}
    \widetilde{\alpha}^{00} \\
    \widetilde{\alpha}^{01} \\
    \widetilde{\alpha}^{02} \\
    \widetilde{\alpha}^{03} \\
    \widetilde{\alpha}^{11} \\
    \widetilde{\alpha}^{12} \\
    \widetilde{\alpha}^{13} \\
    \widetilde{\alpha}^{22} \\
    \widetilde{\alpha}^{23} \\
  \end{array}
\right)
 \equiv
\left(
  \begin{array}{c}
    (4/3)\,\overline{\kappa}^{00}\\
    2\,\delta\widetilde{\kappa}^{01} \\
    2\,\delta\widetilde{\kappa}^{02} \\
    2\,\delta\widetilde{\kappa}^{03} \\
    \delta\widetilde{\kappa}^{11} \\
    \delta\widetilde{\kappa}^{12} \\
    \delta\widetilde{\kappa}^{13} \\
    \delta\widetilde{\kappa}^{22} \\
    \delta\widetilde{\kappa}^{23} \\
  \end{array}
\right),
\end{equation}
where $\vec{\alpha}$ denotes the corresponding vector in the
parameter space $\mathbb{R}^9$ with squared Euclidean norm
\begin{equation}\label{eq:alpha-space-norm}
|\vec{\alpha}|^2 \equiv \sum_{l=0}^{8}\:\left(\alpha^l\,\right)^2\,.
\end{equation}

At this moment, it can also be mentioned that,
for \modMaxth, the phase velocity is given by $\vecbf{v}_\text{phase}
\equiv \widehat{\vecbf{k}}\,\omega/|\vecbf{k}| =
c\,\widehat{\vecbf{k}}\,\big(1-\Xi(\widehat{\vecbf{k}})\big)$,
with a dimensionless function $\Xi$ depending on the wave vector
direction $\widehat{\vecbf{k}}$
and the components of the $\kappa$--tensor \cite{KaufholdKlinkhamer2007}.
This result implies that, for \modMaxth, the phase velocity
$\vecbf{v}_\text{phase}$ and the front velocity
$\vecbf{v}_\text{front}\equiv
\lim_{|\vecbf{k}|\to\infty} \vecbf{v}_\text{phase}(\vecbf{k})$
are equal for each direction $\widehat{\vecbf{k}}$.
Moreover, the phase and group velocities are equal to leading
order in the Lorentz-violating parameters,
$\vecbf{v}_\text{group}\equiv  \partial\omega/\partial\vecbf{k}=
\vecbf{v}_\text{phase}+\BigO(\widetilde{\kappa}^2)$.
Recall that the front velocity
is the relevant quantity for signal propagation and causality
and that the group velocity at the dominant frequency component
of a broad wave packet gives the velocity of energy transport;
see, e.g., Refs.~\cite{Brillouin1960,Jackson1975}.

We now add spin--$\textstyle{\frac{1}{2}}$ particles
with electric charge $e$
and mass $M$ to the theory, taking the usual minimal coupling
to the gauge field \cite{ItzyksonZuber1980}.
The action of these charged particles is assumed to be Lorentz invariant,
so that, for the particular spacetime coordinates employed,
the Lorentz violation of the combined theory  resides solely
in the $\kappa$ term of \eqref{eq:modM-action}. Specifically,
the action with the standard Dirac term included is given by
\begin{align}\label{eq:modMstandD-action}
&\action_\text{modM+standD} = \action_\text{modM}
+\int_{\mathbb{R}^4} \id^4 x \;
\overline\psi(x)
\Big( \gamma^\mu \big(\ii\,\partial_\mu -e A_\mu(x) \big) -M\Big)
\psi(x)\,.
\end{align}
Note that the maximal attainable velocity of this
charged particle equals $c$, which may or may not exceed the
phase velocity of light discussed in the previous paragraph.
See Refs.~\cite{AdamKlinkhamer2001,KosteleckyLehnert2001,BrosEpstein2002}
for further discussion on microcausality in Lorentz-violating theories
and, in particular, Eq.~(50) of Ref.~\cite{KosteleckyLehnert2001}
for microcausality in the fermionic $c_{00}$--model,
which is formally related to a special case
of \modMaxth~\eqref{eq:modMstandD-action} by an appropriate linear coordinate
transformation \cite{BaileyKostelecky2004,Altschul2007PRL98}.

As mentioned in the Introduction, new phenomena at the
energy scale $E_\text{Planck}\approx 10^{19}\,\text{GeV}$
may lead to Lorentz violation in the low-energy theory,
possibly described by \modMaxth~\eqref{eq:modM-action}.
The crucial point to realize is that, \emph{a priori},
the Lorentz-violating parameters $\kappa^{\mu\nu\rho\sigma}$
need not be small (e.g., suppressed by inverse powers of
$E_\text{Planck}$) but can be of order unity
\cite{Collins-etal2004,BernadotteKlinkhamer2007,Klinkhamer2007}.
For this reason, it is of importance to obtain as
strong bounds as possible on \emph{all}
deformation parameters $\kappa^{\mu\nu\rho\sigma}$.

\subsection{Cherenkov threshold condition and UHECR bounds}
\label{sec:cherenkov-threshold-condition}

Vacuum Cherenkov radiation for modified quantum electrodynamics
\eqref{eq:modMstandD-action} with the photonic action
\eqref{eq:modM-action}--\eqref{eq:nonbirefringent-ansatz}
has been studied in the classical approximation
by Altschul \cite{Altschul2007PRL98}
and at tree level by Kaufhold and Klinkhamer \cite{KaufholdKlinkhamer2007}.
As explained in Ref.~\cite{KaufholdKlinkhamer2007}
(see also Ref.~\cite{ColemanGlashow1997}), the radiated energy
rate of a primary particle with point charge $Z_\text{prim}\,e$,
mass $M_\text{prim}> 0$, momentum $\boldsymbol{q}_\text{prim}$,
and ultrarelativistic energy
$E_\text{prim} \sim c\,|\boldsymbol{q}_\text{prim}|$
is asymptotically given by
\begin{equation}\label{eq:dWdt}
\frac{\dd W_\text{modM}(\boldsymbol{q}_\text{prim})}{\dd t}\;
\sim \; Z_\text{prim}^2\; \frac{e^2}{4\pi}\;
\xi(\widehat{\vecbf q}_\text{prim})\;E_\text{prim}^2/\hbar\,,
\end{equation}
where the nonnegative dimensionless coefficient $\xi$
results from appropriate contractions
of the \mbox{$\kappa$--tensor} \eqref{eq:nonbirefringent-ansatz}
with two rescaled \mbox{$q$--vectors}.
The coefficient $\xi$ may or may not depend on the primary particle
direction $\widehat{\vecbf q}_\text{prim}$.

The asymptotic behavior shown in \eqref{eq:dWdt}
holds only for particle energies $E_\text{prim}$ well above the
Cherenkov threshold, which has the following order of magnitude:
\begin{equation}\label{eq:Ethreshold}
E_\text{thresh} \sim M_\text{prim}\, c^2/\sqrt{\widetilde{\kappa}}\,,
\end{equation}
for an appropriate scale $\widetilde{\kappa}$
obtained from the $\widetilde{\kappa}^{\mu\nu}$
components written in terms of the parameters $\alpha^l$
(the scale $\widetilde{\kappa}$ is effectively set to zero if Cherenkov
radiation is not allowed). For completeness, we mention that, at present,
the most detailed study of vacuum Cherenkov radiation
has been performed for another Lorentz-violating theory,
Maxwell--Chern--Simons  theory, and refer
the reader to Refs.~\cite{LehnertPotting2004PRL93,LehnertPotting2004PRD70,
KaufholdKlinkhamer2006,Altschul2007PRD75,KaufholdKlinkhamer2007}
for further discussion and references.

Continuing with nonbirefringent \modMaxth, it is now possible to derive
an upper bound on a particular combination of the deformation parameters
$\alpha^l$ from the observation of a single UHECR event
with a nuclear primary moving in the direction $\widehat{\vecbf q}_\text{prim}$
and having an ultrarelativistic energy $E_\text{prim} \gg M_\text{prim}\,c^2$.
The argument is remarkably simple \cite{Beall1970,ColemanGlashow1997}:
an UHECR proton or nucleus
can arrive on Earth only if it does not lose energy by vacuum
Cherenkov radiation and this requires the particle energy $E_\text{prim}$
to be at or below threshold, $E_\text{prim} \leq E_\text{thresh}$.

The \emph{caveat} of this simple argument is
that the radiation rate \eqref{eq:dWdt} should not be
suppressed by an extremely small numerical factor entering the
coefficient $\xi$. But there is no reason to expect the
presence of such an extremely small numerical factor;
cf. Refs.~\cite{ColemanGlashow1997,KaufholdKlinkhamer2007}.
In fact, a recent calculation \cite{KaufholdKlinkhamerSchreck2007}
of the tree-level spinor-particle radiation rate for
the special case of having only a single nonzero parameter $\alpha^0 >0$
(i.e., case 2 of App.~C of Ref.~\cite{KaufholdKlinkhamer2007}) gives
coefficient $\xi(\widehat{\vecbf q}_\text{prim})=(7/24)\,\alpha^0$
on the right-hand side of \eqref{eq:dWdt}.
With $\xi$  of order $\widetilde{\kappa}>0$,
the particle would slow down from an initial
energy $E_\text{prim}\gg E_\text{thresh}$ to the threshold energy
$E_\text{thresh}$ after having traveled a distance of the order of meters
rather than parsecs (at least, for the values of $\widetilde{\kappa}$
and $E_\text{prim}$ considered in this article).

It may be worthwhile to repeat that
expression \eqref{eq:dWdt} holds
for an electric point charge, whereas we expect a form factor
for a finite charge distribution with typical length scale
$a \equiv \hbar\,c/\Lambda$; see also the discussion in
Refs.~\cite{Afanasiev2004,Altschul2007hepth}.
But this modification of the theory
would primarily affect the total radiation rate and not the energy
threshold \eqref{eq:Ethreshold} on which our bounds are based, at least,
for large enough cutoff $\Lambda$.
Anyway, the theory considered in this article is precisely the one given
in Sec.~\ref{sec:nonbirefringent-modified-Maxwell-theory},
without additional contact terms.

Using the explicit (direction-dependent) result
\cite{Altschul2007PRL98} for the threshold energy \eqref{eq:Ethreshold},
the condition $E_\text{prim} \leq E_\text{thresh}$ can then be written as
the following upper bound on the deformation parameters
\eqref{eq:alpha-parameters} of \modMaxth~\eqref{eq:modMstandD-action}:
\begin{equation}\label{eq:generalbound}
R\big(\,
\alpha^{0}+\alpha^{j}\: \widehat{\vecbf q}_\text{prim}^{j}
+\widetilde{\alpha}^{jk}\:
 \widehat{\vecbf q}_\text{prim}^{j}\,\widehat{\vecbf q}_\text{prim}^{k}
\big)
\leq
\big(M_\text{prim}\,c^2/E_\text{prim} \big)^2 ,
\end{equation}
with each index $j$ and $k$ summed over 1 to 3
and ramp function $R(x) \equiv (x + |x|\,)/2$.
The parameters $\widetilde{\alpha}^{jk}$ appearing in the argument of
the ramp function on the left-hand side of \eqref{eq:generalbound} are
defined in \eqref{eq:alpha-parameters}, with
$\widetilde{\alpha}^{33}\equiv-\widetilde{\alpha}^{11}-\widetilde{\alpha}^{22}$
from the tracelessness condition \eqref{eq:widetilde-kappa}.

Observe that, as expected on general grounds,
bound \eqref{eq:generalbound} is invariant under a simultaneous
rotation of $\widehat{\vecbf q}_\text{prim}$ and appropriate redefinition of
parameters $\alpha^{1} \cdots \alpha^{8}$.
This becomes especially clear if the argument of the ramp function
on the left-hand side of \eqref{eq:generalbound} is written as
$\alpha^{0}+(\boldsymbol{\alpha}\cdot\widehat{\vecbf q}_\text{prim})
+(\boldsymbol{\beta} \cdot\widehat{\vecbf q}_\text{prim})\,
 (\boldsymbol{\gamma}\cdot\widehat{\vecbf q}_\text{prim})$,
with two orthogonal 3--vectors $\boldsymbol{\beta}$ and $\boldsymbol{\gamma}$
replacing the 3--tensor $\widetilde{\alpha}^{jk}\,$.
In the same way, it is possible to absorb
a parity-reflection of $\widehat{\vecbf q}_\text{prim}$
by a change of sign of the three parameters entering $\boldsymbol{\alpha}$,
while leaving the other six parameters unchanged.

In the following, we will use \eqref{eq:generalbound}
to obtain bounds on all nine deformation parameters $\alpha^l$
from a sufficiently large number $N$ of UHECR events
distributed over a large enough part of the sky
(having primary energies $E_n$ and directions $\widehat{\vecbf q}_n$,
for $n=1, \ldots, N$) and an appropriate value of $M_\text{prim}$.

\section{Data selection}
\label{sec:data-selection}

The Pierre Auger Collaboration has published a table of 29 UHECR events
with energies above $10\,\text{EeV}\equiv 10^{19}\,\text{eV}$
\cite{Abraham-etal2006}.
In addition, the atmospheric depth of the shower maximum, $X_{\rm max}$,
was determined for all of these events. Since the presently available
Auger results indicate a mixed composition above 10~EeV~\cite{Unger-etal2007}
and our upper bound \eqref{eq:generalbound} scales with $M_\text{prim}^2$,
we aim at selecting events initiated by \emph{light} primaries
primaries such as protons or helium nuclei. For this reason,
we apply the following selection criteria to the list of 29 events
from Ref.~\cite{Abraham-etal2006}:
\begin{itemize}
\item[(1)]
$X_{\rm max}/\big(\text{g}\;\text{cm}^{-2}\big)
\geq 750+50\:\big(\log_{10}\,(E/\text{EeV}) -1\big)$;
\item[(2)]
$\Delta_\gamma \geq +2$, where $\Delta_\gamma$ quantifies, in units
     of standard deviation, the difference between the observed
     $X_{\rm max}$ and the average $X^{\gamma}_{\rm max}$ expected
     from primary photons [\,see Eq.~(1) of Ref.~\cite{Abraham-etal2006}
     for the definition of $\Delta_\gamma$\,];
\item[(3)]
primary energy between 10 and 30 EeV.
\end{itemize}

\begin{table*}
\begin{center}
\caption{Simulated average shower-maximum atmospheric depth
$X_{\rm max}$  in units of $\text{g}\:\text{cm}^{-2}$, with
root-mean-square in brackets. Uncertainties are of
the order of 3~$\text{g}\:\text{cm}^{-2}$ for nuclear primaries and
5~$\text{g}\:\text{cm}^{-2}$ for photons. The calculations were performed with
\textsc{CORSIKA}~\cite{CORSIKA1998}
using  two different hadronic interaction models,
\textsc{QGSJET}~01 (abbreviated Q)~\cite{QGSJET1997} and \textsc{SIBYLL}~2.1
(abbreviated S)~\cite{SIBYLL1999}.
The asterisk ($\ast$) indicates that, for primary photons above
10$^{19.5}$~eV, the $X_{\rm max}$ values also depend on the direction
of the event due to geomagnetic cascading (see, e.g.,
Ref.~\cite{Risse-Homola2007} and references therein).
For instance, the $X_{\rm max}$ values
at $10^{20}$~eV range between 940(85) and 1225(175)
$\text{g}\:\text{cm}^{-2}$,
according to Ref.~\cite{Homola-etal2005}.
\vspace*{.5\baselineskip}}
\label{tab-xmax}
\renewcommand{\tabcolsep}{0.5pc}    
\renewcommand{\arraystretch}{1.1}   
\begin{tabular}{lcccccccc}
\hline\hline
$\log_{10}(E/\text{eV})$
& 17.5  & 18.0 &  18.5 &  19.0  & 19.5  & 20.0 &
20.5  & 21.0 \\
\hline
Q---p  &  695(65) & 725(65) & 755(65) & 775(65) & 800(60) & 820(60) &
845(55) & 865(55) \\
Q---He & 665(45)& 697(45) &725(45) &750(45) &775(40) &800(40) &825(38)
&850(35) \\
Q---O  &         &     &  693(31) &717(30) &745(30) &780(30) &800(28)& \\
Q---Fe & 600(25) &632(22) &664(22) &695(22) &725(20) &755(20) &782(20)
&810(20) \\
\hline
S---p  &      &  740(65) &       & 800(60) &      &  860(55) &885(50) &\\
S---Fe   &      & 640(22)  &      & 700(22)  &     &  755(20) &785(20) &\\
\hline
photon  &   &    &       915(60) &975(70) &1075(95)
&($\ast$)&($\ast$)&($\ast$)
\\
\hline\hline
\end{tabular}
\end{center}
\end{table*}

The first (and most selective) criterion aims at rejecting heavier
nuclei. The chosen $X_{\rm max}$ parametrization roughly follows the
average $X_{\rm max}$ values expected for helium nuclei. Thus, about
half of the number of helium nuclei are rejected by this cut, while keeping
the majority of protons. Oxygen and iron nuclei are rejected at a high level.
In terms of the parameter $\Delta_\text{prim}$,
the selected events roughly have
$\Delta_{\rm oxygen} \leq -1$ and $\Delta_{\rm iron} \leq -2$,
respectively. Table~\ref{tab-xmax} gives some calculated values
for $X_{\rm max}$ as a function of
primary energy and primary type (hydrogen, helium, oxygen, and iron nuclei
and photon).

\begin{table}
\begin{center}
\caption{Selected Auger events from Ref.~\cite{Abraham-etal2006}:
event identification number, primary energy $E$ [EeV],
shower-maximum atmospheric depth $X_{\rm max}$ [$\text{g}\:\text{cm}^{-2}$],
and pseudo-random event directions with
right ascension $\text{RA}'\in [0^{\circ},360^{\circ}]$
and declination $\delta'\in [-70^{\circ},25^{\circ}]$.
The primes on the ID numbers are to emphasize the nonreality
of these event directions, which can later be replaced by the
measured values RA and $\delta$
after their release by the Pierre Auger Collaboration
(dropping the primes on the ID Nos.).\vspace*{.5\baselineskip}}
\label{tab-Auger-events}
\renewcommand{\tabcolsep}{1.5pc}    
\renewcommand{\arraystretch}{1.1}   
\begin{tabular}{lccc}           
\hline\hline
 ID No. & $E$ & $X_{\rm max}$  &(RA$'$, $\delta'$)\\
\hline
 668949$'$    & 18 & 765 & $(356 , -29 )$\\
 673409$'$    & 13 & 760 & $(344 , -62 )$\\
 828057$'$    & 14 & 805 & $(086 , -34 )$\\
 986990$'$    & 16 & 810 & $(152 , -33 )$\\
1109855$'$    & 17 & 819 & $(280 , -30 )$\\
1171225$'$    & 16 & 786 & $(309 , -70 )$\\
1175036$'$    & 18 & 780 & $(228 , +17 )$\\
1421093$'$    & 27 & 831 & $(079 , +13 )$\\
1535139$'$    & 16 & 768 & $(006 , -62 )$\\
1539432$'$    & 13 & 787 & $(153 , -15 )$\\
1671524$'$    & 14 & 806 & $(028 , -63 )$\\
1683620$'$    & 21 & 824 & $(024 , -23 )$\\
1687849$'$    & 17 & 780 & $(031 , -23 )$\\
2035613$'$    & 12 & 802 & $(079 , -08 )$\\
2036381$'$    & 29 & 782 & $(158 , -03 )$\\
\hline\hline
\end{tabular}
\end{center}
\vspace*{0cm}
\end{table}

The second criterion is to ensure no contamination from primary photons,
as neutral photons would not emit vacuum Cherenkov radiation at tree level.
In fact, all 29 events from the original table of
Ref.~\cite{Abraham-etal2006} already have $\Delta_\gamma \geq +2$.

The third criterion provides for a more
homogeneous sample, but can certainly be relaxed in a future analysis.
We have also accounted for the increased missing-energy correction in case of
a nuclear primary, that is, the energies given in Ref.~\cite{Abraham-etal2006}
(which refer to photon primaries) were increased by
$7\,\%$,  according to Ref.~\cite{Pierog-etal2006}.

In total, 15 events remain which are listed in Table~\ref{tab-Auger-events}.
The corresponding event directions have not yet been published
by the Pierre Auger Collaboration.
Shown in Table~\ref{tab-Auger-events} are
pseudo-random directions chosen from uniform distributions
of right ascension $\text{RA}'\in [0^{\circ},360^{\circ}]$ and
declination $\delta'\in [-70^{\circ},25^{\circ}]$, where
the primes indicate the fictional nature of these directions.
Note that the presently known UHECRs at energies $E \sim 10\,\text{EeV}$
are consistent with the hypothesis of large-scale isotropy; see, e.g.,
Refs.~\cite{BhattacharjeeSigl1998,Stanev2004,Takeda1999,Anchordoqui-etal2003,
Abbasi-etal2004,Mollerach2007,Armengaud2007} for a selection of the available data.

All bounds of this paper are based on the values given
in Table~\ref{tab-Auger-events},
but, for the reason given above, we expect that
the same bounds are obtained from the actual event directions
when they are made available by the Pierre Auger Collaboration.
Just to be clear,  the bounds of this article simply follow from
having 15 UHECR events with more or less equal energies
of order $10$~EeV and more or less random directions
over a significant part of the sky, the precise association of energy
and direction being irrelevant.

The average energy of the sample given in Table~\ref{tab-Auger-events}
is $\langle E \rangle \approx 17$~EeV with
an uncertainty of about 25\,\%~\cite{Bellido-etal2005}.
A larger uncertainty concerns the identity of the primary particle.
We estimate an average mass for the selected events of
$\langle M_\text{prim}\rangle \approx f\times 5$~GeV/c$^{2}$,
where the factor $f \in [0.2, 1.8]$
indicates the uncertainty ($f = 0.2$ refers to a pure proton beam,
$f = 1$ to a mixed four-component primary composition which fits the Auger
data \cite{Unger-etal2007}, and $f = 1.8$ to the extreme case of no
protons at all in the primary beam and a mixed three-component composition).
In the next section, we will use an even larger value of $f$,
namely $f=3.2$, in order to obtain absolutely reliable bounds
on the deformation parameters $\alpha^l$.
%
%
%
%

\section{UHECR Cherenkov bounds}
\label{sec:UHECRbounds}

Following up on the remarks of
the last paragraph of Sec.~\ref{sec:cherenkov-threshold-condition},
it is now possible to determine numerically an exclusion
domain in the $\vec{\alpha}$ parameter space $\mathbb{R}^9$
from the $N=15$ events of Table~\ref{tab-Auger-events} and
the Cherenkov threshold condition \eqref{eq:generalbound}.
This exclusion procedure only works for those parameters $\vec{\alpha}$
for which the phase velocity of light is strictly less than the maximal
attainable velocity $c$ of the charged particle in
theory \eqref{eq:modMstandD-action}, $v_\text{phase}<c$.
In fact, this phase-velocity condition
corresponds to having a positive argument of the ramp
function on the left-hand side of \eqref{eq:generalbound}.
Hence, the considered domain of parameter space is given by
\begin{equation}\label{eq:Dcausalopen}
D_\text{causal}^\text{(open)}
\equiv
\{ \vec{\alpha} \in \mathbb{R}^9 :\;
\forall_{\widehat{x}\in \mathbb{R}^3} \;
(\alpha^0 + \alpha^j\,\widehat{x}^j+
\widetilde{\alpha}^{jk}\,\widehat{x}^j\,\widehat{x}^k)>0 \},
\end{equation}
with $\widehat{x}$ an arbitrary unit vector in Euclidean 3--space,
indices $j$ and $k$ summed over 1 to 3, and
parameters $\widetilde{\alpha}^{jk}$ defined by \eqref{eq:alpha-parameters}.
As mentioned in Ref.~\cite{KaufholdKlinkhamer2007} and
Sec.~\ref{sec:nonbirefringent-modified-Maxwell-theory}, the condition
$v_\text{front} =v_\text{phase} \leq c$ is, most likely,
necessary for having a consistent and causal version of \modMaxth,
hence the suffix ``causal'' in \eqref{eq:Dcausalopen}.
The other suffix ``(open)'' in \eqref{eq:Dcausalopen}
denotes the restriction of the causal domain
by use of the open inequality ``$(\alpha^0 +\cdots)> 0$''
instead of the closed inequality ``$(\alpha^0 +\cdots)\geq 0$'',
because no Cherenkov limits can be obtained for $v_\text{phase} = c$.

A hypersphere $S^8_a$ of finite radius $a$ in the subspace
\eqref{eq:Dcausalopen} can be found numerically,
so that for \emph{each}
$\vec{\alpha} \in S^8_a$ the inequality \eqref{eq:generalbound}
is violated for \emph{at least} one event from
Table~\ref{tab-Auger-events}. The excluded domain of parameter space then
corresponds to the region on or outside this hypersphere.
Observe, namely, that the left-hand side of \eqref{eq:generalbound},
for a positive argument of the ramp function,
increases by a factor $\lambda$
under the scaling of $\vec{\alpha} \to \lambda\,\vec{\alpha}$ with $\lambda>1$,
so that inequality \eqref{eq:generalbound} is violated
for $\lambda>1$ if it is already for $\lambda=1$.

In this way, the experimental data of Table~\ref{tab-Auger-events}
allow us to exclude the following region of parameter space
at the two--$\sigma$ level:
\begin{subequations}\label{eq:Dexcluded}
\begin{eqnarray}
D_\text{excluded}^a &=& D_\text{causal}^\text{(open)} \cap
\{ \vec{\alpha} \in \mathbb{R}^9 :\;
|\vec{\alpha}| \geq a \},
\label{eq:Dexcluded-general}\\[3mm]
a &\approx& 3 \times 10^{-18}\,
            \left(\frac{M_\text{prim}}{16\:\text{GeV}/c^2}\right)^{2},
\label{eq:Dexcluded-avalue}
\end{eqnarray}
\end{subequations}
where $|\vec{\alpha}|$ is the standard Euclidean
norm \eqref{eq:alpha-space-norm} and where the reference value for the
mass of the primary charged particle has conservatively
been taken equal to that of oxygen (as mentioned in
Sec.~\ref{sec:data-selection}, most primaries
in the sample are expected to be protons and helium nuclei).
The $2\sigma$ statistical
error included in the number $3 \times 10^{-18}$ on the right-hand side
of \eqref{eq:Dexcluded-avalue} is mainly
due to the $25\,\%$ energy uncertainty of the UHECRs.
%
%
%
%

The domain \eqref{eq:Dexcluded} gives only part of the excluded region,
because at certain points on the hypersphere $S^8_a$ the excluded region
pushes in. An interesting question, then, is how far the excluded region
extends inwards. A partial answer is given by the following result.
It is possible to establish numerically the existence of a
hyperball $B^9_b$ with finite radius $b$, so that for \emph{each}
$\vec{\alpha} \in B^9_b$ the inequality \eqref{eq:generalbound}
holds for \emph{all} events from Table~\ref{tab-Auger-events}.
Specifically, the following domain of parameter space is found to be allowed:
\begin{subequations}\label{eq:Dallowed}
\begin{eqnarray}
D_\text{allowed}^b &=& D_\text{causal} \cap
\{ \vec{\alpha} \in \mathbb{R}^9 :\;
|\vec{\alpha}| \leq b \},
\label{eq:Dallowed-general}\\[3mm]
b &\approx&  (1.5 \pm 0.7) \times 10^{-19}
             \,\left(\frac{M_\text{prim}}{16\:\text{GeV}/c^2}\right)^{2},
\label{eq:Dexcluded-bvalue}
\end{eqnarray}
\end{subequations}
with the approximate $1\sigma$ error in \eqref{eq:Dexcluded-bvalue}
being mainly due to the energy uncertainty of the UHECRs
and using the same reference value for the mass $M_\text{prim}$
as in \eqref{eq:Dexcluded-avalue}.
The complete allowed region may very well extend beyond the radius
$b$, but not beyond the radius $a$ as given by \eqref{eq:Dexcluded-avalue}.

Returning to the exclusion domain \eqref{eq:Dexcluded},
the corresponding $2\sigma$ bound on $|\vec\alpha|$ is given by
\begin{widetext}  
\begin{equation}
\vec{\alpha} \in D_\text{causal}^\text{(open)}:\;
\big|\,\vec{\alpha}\,\big|^2 \equiv
\sum_{l=0}^{8}\:\big(\alpha^l\,\big)^2  < \big(\, 3 \times 10^{-18}\,\big)^2
\,\left(\frac{M_\text{prim}}{16\:\text{GeV}/c^2}\right)^{4},
\label{eq:alpha-Cherenkov-bound}
\end{equation}
\end{widetext}
where, as argued above, a value of $16\:\text{GeV}/c^2$ for
$M_\text{prim}$ is entirely reasonable.
With Lorentz invariance violated by the \modMaxth,
it is important to specify the reference frame in which bound
\eqref{eq:alpha-Cherenkov-bound} holds, namely,
the solar-system frame in which the cosmic-ray energies are measured.

The new bound \eqref{eq:alpha-Cherenkov-bound}
is consistent with earlier expectations, as given by
the one-sided bound (C12) of Ref.~\cite{KaufholdKlinkhamer2007}
in a somewhat implicit  notation.
Remark that our new bound on $\vec{\alpha}$ is also
``one-sided'' (vacuum Cherenkov radiation occurs only
if the light velocity is less than that of the charged particle),
but, now, the one-sidedness is in a 9--dimensional space.
In order to be as clear as possible,
the exclusion domain \eqref{eq:Dexcluded} has explicitly
been given as an overlap with the domain \eqref{eq:Dcausalopen}
defined at the beginning of this section
and bound \eqref{eq:alpha-Cherenkov-bound} has been
specified to hold for $\vec{\alpha}$ in the same domain.

The domains \eqref{eq:Dexcluded} and \eqref{eq:Dallowed},
together with the resulting upper bound \eqref{eq:alpha-Cherenkov-bound},
have been calculated from the measured (unprimed) energies and
pseudo-random (primed) directions shown in Table~\ref{tab-Auger-events}.
But, as mentioned in Sec.~\ref{sec:data-selection},
we expect that the same results are obtained if the measured
(unprimed) event directions are used.

In closing, it may be helpful to rephrase our main result
\eqref{eq:alpha-Cherenkov-bound} as follows:
considering deformation parameters $\alpha^0, \dots, \alpha^8$
with a corresponding phase velocity of light less than the maximal
attainable velocity of charged particles, each of these nine
parameters must separately have a modulus less than the value $a$ given
by \eqref{eq:Dexcluded-avalue} at the $2\sigma$ level.
A more detailed analysis of these bounds is
left to the future when more UHECR events become available.

\section{Summary}
\label{sec:summary}

In this article, we have established an UHECR Cherenkov
bound \eqref{eq:alpha-Cherenkov-bound} on the nine deformation
parameters \eqref{eq:alpha-parameters} of nonbirefringent
modified-Maxwell theory \eqref{eq:modM-action}--\eqref{eq:nonbirefringent-ansatz}
coupled to a standard Dirac particle,
with total action \eqref{eq:modMstandD-action}.
Three remarks may be helpful to clarify the
background and meaning of our result.

First, the derivation of bound \eqref{eq:alpha-Cherenkov-bound} relies,
strictly speaking, on the assumption of large-scale isotropy
of the arrival directions of 10--30 EeV cosmic rays, which is,
however, supported by all the available data
(see Sec.~\ref{sec:data-selection} for further discussion).
Second, this bound is essentially ``one-sided'' as it
applies only to parameters belonging to
domain \eqref{eq:Dcausalopen}, which may, however, correspond
to a large part of the ``physical domain'' of the theory
(see Secs.~ \ref{sec:nonbirefringent-modified-Maxwell-theory}
and \ref{sec:UHECRbounds} for further discussion).
Third, bound \eqref{eq:alpha-Cherenkov-bound} holds for the theory
as defined by the action \eqref{eq:modMstandD-action}
but also applies to a theory with additional Lorentz-violating
parameters $c^{\mu\nu}$ in the Dirac sector
\cite{ColladayKostelecky1998},
provided the parameters $\widetilde{\kappa}^{\mu\nu}$
in bound \eqref{eq:alpha-Cherenkov-bound} are replaced by
the effective parameters $\widetilde{\kappa}_\text{eff}^{\mu\nu}\equiv
\widetilde{\kappa}^{\mu\nu}-2\, c^{\mu\nu}$
\cite{BaileyKostelecky2004,Altschul2007PRL98,Altschul2007hepth}.

Having clarified the background of our Cherenkov
bound \eqref{eq:alpha-Cherenkov-bound}, it is to be remarked that
the corresponding bound at the $10^{-18}$ level
for the spatially isotropic nonbirefringent
deformation parameter $\alpha^0$
improves significantly upon the direct laboratory bound at the $10^{-7}$
level [6(c)]  
or the indirect (electron anomalous-magnetic-moment)
laboratory bound at the $10^{-8}$ level [7(c)]. 
But also the other Cherenkov bounds at the $10^{-18}$ level
for the nonisotropic  parameters $\alpha^l$, for $l=1, \ldots, 8$,
improve considerably upon the direct laboratory bounds at the
$10^{-12}$ to $10^{-16}$ levels \cite{Stanwix-etal2006}.
While we note that a consistent comparison of bounds from laboratory
and astrophysics data needs to be performed with great care
(for example, Ref.~[8(b)]  
allows for independent Lorentz-violating
parameters in the photonic and fermionic sectors),
the astrophysics bounds at the $10^{-18}$ level are certainly indicative.

With more and more UHECR events detected at the Pierre Auger Observatory
over the coming years, these Cherenkov bounds at the $10^{-18}$ level
can be expected to improve on
at least two grounds. First, the availability of a larger number of events
and further observables in addition to $X_\text{max}$
will allow for more stringent cuts to obtain an essentially
pure proton sample, which would give a reduction factor of $16^2$
in \eqref{eq:Dexcluded-avalue}.\footnote{If a sufficient number of
10 EeV cosmic-ray electrons could somehow be identified, there
could be a further reduction factor in \eqref{eq:Dexcluded-avalue} of
approximately $(938/0.511)^2$ $\approx$ $3 \times 10^{6}$
compared to the proton case.}
Second, higher energy events will
become available for the analysis, perhaps up to values of
order $E \sim 3 \times 10^{20} \, \text{eV}$
(the current highest known energy \cite{Bird-etal1995}),
which would give a further reduction factor of approximately $10^3$
in \eqref{eq:Dexcluded-avalue}.\footnote{The
steeply falling flux spectrum and the expected suppression
of protons from cosmologically distant sources with energies
above $6 \times 10^{19}\,\text{eV}$ \cite{GZK1966,Roth2007} present
an experimental challenge for pushing the energy to substantially higher values.}
Combined, the Cherenkov bounds on the nine deformation
parameters $\alpha^l$ can perhaps be reduced by a
factor of order $10^5$ to the $10^{-23}$ level in the next decade.

\section*{\hspace*{-4.5mm}ACKNOWLEDGMENTS}

It is a pleasure to thank C. Kaufhold and D. K\"{u}mpel for discussions
and technical support.

\section*{\hspace*{-4.5mm}NOTE ADDED IN PROOF}

The Pierre Auger Collaboration has recently published a list of 27 UHECR
events with energies above $57\;\text{EeV}$ and accurate event directions,
which were recorded over the period 1 January 2004 to 31 August 2007
\cite{Abraham-etal2007}.
In order to provide better coverage of the northern celestial hemisphere,
we add a $210\;\text{EeV}$ event from the AGASA array
\cite{Hayashida-etal1994}
and the previously mentioned $320\;\text{EeV}$ event from the Fly's Eye detector
\cite{Bird-etal1995},
both with reasonably accurate arrival directions. The observed energies and
directions of these 29 events can now be used to sharpen bound
\eqref{eq:alpha-Cherenkov-bound}
in terms of the square of the number $a$ given by
\eqref{eq:Dexcluded-avalue}.

In the absence of specific information on the chemical composition
of all 29 events, we simply take $M_\text{prim} \approx 16\;\text{GeV}/c^{2}$.
From these particular 29 UHECR events with an estimated energy uncertainty of
the order of $25\,\%$, the improved value for $a$ is found to be given by
\begin{equation}
A \equiv a_\text{new}
\approx 2 \times 10^{-19}\,
\left(\frac{M_\text{prim}}{16\:\text{GeV}/c^2}\right)^{2},
\label{eq:Dexcluded-avalue-new}
\end{equation}
which gives the $2\sigma$ excluded region $D_\text{excluded}^{A}$ as defined by
\ref{eq:Dexcluded-general}.
With the 29 event directions known explicitly,
there is no assumption entering the corresponding $2\sigma$ bound
\eqref{eq:alpha-Cherenkov-bound}
with $A^2$ on the right-hand side, apart from the uncertainty in the primary
mass and the \emph{caveat} mentioned in
Sec.~\ref{sec:cherenkov-threshold-condition}.

\newpage

\end{document}